# High-energy-density plasma in femtosecond-laser-irradiated nanowire array targets for nuclear reactions

**Defeng Kong[1], Guoqiang Zhang[2,*], Yinren Shou[1], Shirui Xu[1], Zhusong Mei[1], Zhengxuan Cao[1], Zhuo Pan[1], Pengjie Wang[1], Guijun Qi[1], Yao Lou[2], Zhiguo Ma[3], Haoyang Lan[1,4], Wenzhao Wang[5], Yunhui Li[6], Peter Rubovic[7,8], Martin Veselsky[7], Aldo Bonasera[9,10], Jiarui Zhao[1], Yixing Geng[1], Yanying Zhao[1], Changbo Fu[3], Wen Luo[4], Yugang Ma[2,3], Xueqing Yan[1,11,12], and Wenjun Ma[1,11,12,§]**

[1]*State Key Laboratory of Nuclear Physics and Technology, and Key Laboratory of HEDP of the Ministry of Education, CAPT, School of Physics, Peking University, Beijing, 100871, China*

[2]*Shanghai Institute of Applied Physics, Chinese Academy of Sciences, Shanghai, 201800, China*

[3]*Key Laboratory of Nuclear Physics and Ion-Beam Application (MOE), Institute of Modern Physics, Fudan University, Shanghai, 200433, China*

[4]*School of Nuclear Science and Technology, University of South China, Hengyang, 421001, China*

[5]*INPAC and School of Physics and Astronomy, Shanghai Jiao Tong University, Shanghai, 200240, China*

[6]*Shenzhen Institutes of Advanced Technology, Chinese Academy of Sciences, Shenzhen, 518055, China*

[7]*Institute of Experimental and Applied Physics, Czech Technical University in Prague, Husova 240/5, 11000, Prague 1, Czech Republic*

[8]*ELI Beamlines Center, Institute of Physics of the Czech Academy of Sciences, 252 41 Dolní Břežany, Czechia*

[9]*Cyclotron Institute, Texas A&M University, College Station, 77843, TX, United States*

[10]*Laboratori Nazionali del Sud, INFN, via Santa Sofia, 62, Catania, 95123, Italy*

[11]*Beijing Laser Acceleration Innovation Center, Huairou, Beijing, 101400, China*

[12]*Collaborative Innovation Center of Extreme Optics, Shanxi University, Taiyuan, Shanxi 030006, China*

[*]corresponding author: zhangguoqiang@zjlab.org.cn

[§]corresponding author: wenjun.ma@pku.edu.cn

## Abstract

In this work, the high-energy-density plasmas (HEDP) evolved from joule-class-femtosecond-laser-irradiated nanowire array (NWA) targets are numerically and experimentally studied. The particle-in-cell (PIC) simulations indicate that ions accelerated in the sheath field around the nanowires' surface were eventually confined in NWA plasma, contributing most to the high energy densities. The protons emitted from the front surface of NWA targets provide rich information about the interaction. The electron and ion energy densities in a broad target parameter range are given. Compared to planar targets, the ion energy density is one order-of-magnitude higher, and the volume of the HEDP is several-fold larger. At optimal target parameters, 8% of the laser energy can be converted to confined protons and results in ion energy densities of up to GJ/cm$^3$ level. Experimental measurements of the emitted ions and neutrons from $^2$H(d, n)$^3$He fusion from polyethylene or deuterated polyethylene NWA targets confirm the above results.

# 1. Introduction

Far away from the living environment of our human beings, high-energy-density plasma (HEDP) widely exists in the universe, for example, in the cores of planets and stars[1]. Understanding and controlling such a form of matter would help us unveil the origin of heavy elements[2] and harness future nuclear energy[3]. Thanks to the development of high-power laser technology, we can now create HEDP in the lab by compressing and heating a spherical capsule with multiple-MJ ns-lasers. Such matter is under the worldwide focus and expected to reach self-sustainable reaction for inertial confinement fusion(ICF)[4-7], which also promotes the development of laser nuclear physics[8-15], laboratory astrophysical physics[16, 17], and high energy density physics[18-20].

Apart from the well-known ICF scheme, HEDP can be created with ultra-intense and ultrashort laser pulses by virtue of the relativistic laser-plasma interactions. Ultra-intense laser field can deeply ionize the target and accelerate the electrons to MeV levels on femtosecond timescales, forming a strong micron-scale space-charge field driving the ions. The most common experimental scheme is irradiating a solid foil target at ultra-high intensity. The laser pulse interacts with the front surface of the foil, generating supra-thermal electrons in the expanded dilute plasma. Heated by the supra-thermal electrons and the return current, the solid target turns to the HEDP. The problem with this method is that if we are interested in the nuclear reactions in HEDP, the entire solid foil is not hot enough. The indirect heating results in a considerable temperature gradient in the foil[21]. Only a small number of ions close to the surface reach the temperature sufficient for nuclear reactions. In order to overcome the problem, cluster targets produced from cryogenic high-density gas[12-14] or low-density foam targets[22-25] were used. The laser pulse can propagate through them and volumetrically heat the nanometer-sized clusters or the foam's building blocks, resulting in homeothermal HEDP. But the densities of the plasma typically do not exceed one percent of solid density, which is unfavorable for nuclear reactions.

Recently, more and more attention was paid to novel micro-structured targets to acquire near-solid-density HEDP[26-32]. Among them, nanowire array (NWA) targets have shown superior performance. Numerical and experimental studies indicate that NWA targets can absorb the majority of the laser energy and significantly enhance the energy conversion efficiency to plasma compared to solid targets[33-35], producing homeothermal HEDP with near-solid density. Fig.1(a) schematically shows an NWA target. The typical diameter (*D*), length ($L$), and spacing ($S$) are 10s-100s nm, 1-10μm, and 0.1-1μm, respectively. The heating process of NWA targets under ultra-intense laser irradiation is illustrated in fig.1(b). The NWA's structure allows a laser pulse to propagate inside the target and interact with the sidewall of nanowires. Electrons in the skin depth are pulled out and accelerated in the laser field. On top of their quiver motions, they collectively move along the laser propagation direction due to the $v \times B$ force. The resulting return current (typically10s-100s kA)(36, 37) inside nanowires and the bombard of the supra-thermal electrons accelerated by $J \times B$ heating(38) efficiently heat the NWAs to a high temperature. In the meantime, the escaped electrons also give rise to a space-charge field (sheath field) normal to the surface of nanowires. Ions can be subsequently accelerated to MeV-level energy in the sheath field(39, 40), which also contributes to plasma heating when they are stopped by nearby nanowires. More details of the interaction, such as the plasma compression due to the Z-pinch formation(37), are still under exploration.

The emissions of energetic electrons(36, 41), ions(42, 43), x/ γ -rays(44-47) and neutrons(48, 49) from the NWA plasma were extensively observed. Purvis et al. measured the x-rays from Ni and Au NWA targets irradiated with a 0.5 J femtosecond laser pulse(44). 100-fold enhanced line emission over planar solid targets was observed. By virtue of numerical simulations, they concluded that the electron energy density(EED) after nanowires' expansion is 0.3 GJ/cm$^3$, comparable to that of spherical compression at the National Ignition Facility with megajoule laser pulses(50). Later on, they increased the laser energy to 10J(51) and observed an intense line emission from Au$^{72+}$ ions due to the >100 GJ/cm$^3$ EED. Naturally, such a high energy density attracts interest in thermal nuclear reactions. Deuterium-deuterium fusion in NWA was firstly reported in 2018 by using deuterated polyethylene NWA targets(47). It produced a neutron yield of 10$^6$/J/shot, a record for the joule-scale laser. Recently, 10$^7$/shot neutrons were measured at a higher intensity with a PW laser(49).

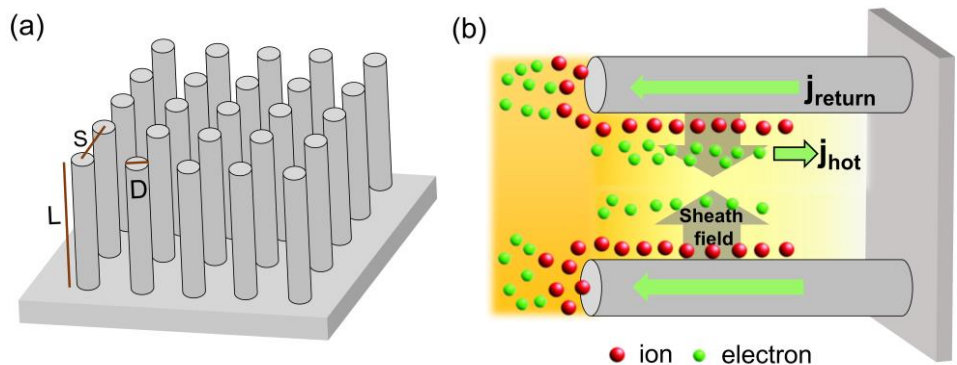

**Figure 1:** (a) Schematic diagram of an NWA target. *L, D,* and *S* mean NWA's length, diameter, and spacing. (b) Plasma heating and ion acceleration picture in an NWA target.

Up to now, the electron heating dynamics and EED in NWA targets have been widely studied by measuring the core-shell x-ray emission from high-Z NWAs. Regrettably, such a method is inapplicable for low-Z NWA targets, which are particularly important for fusion-energy-oriented reactions such as $^2$H(d, n)$^3$He and $^{11}$B(p, 2α)α. Besides, the energy density of ions in NWA targets, which is foremost for nuclear reactions, was rarely discussed. The optimal target parameters for fusion nuclear reactions and more ways to characterize the high-energy-state besides x-ray spectroscopy require further studies.

In this work, we firstly study the interaction between NWA targets and femtosecond laser pulses employing collisional 2D PIC simulation. The electron and ion (proton) energy densities of NWA targets in a broad target parameter range are given and compared to planar targets. The confined and emitted protons are defined. Their interrelationship and contribution to the energy densities of the HEDP are thoroughly discussed. Thereafter, we experimentally measure the emitted protons, deuterium ions, and neutrons from $^2$H(d, n)$^3$He fusion utilizing ordinary/deuterated polyethylene NWA targets, which confirms our simulation results and conclusions.

## 2. Laser-NWA interaction, confined and emitted protons

In order to illustrate the interaction of NWA targets with relativistic femtosecond pulses, 2D particle-in-cell simulations are performed by using the current EPOCH(52) code. The simulation box has a size of $40\lambda_L \times 50\lambda_L$ with $4000 \times 5000$ cells in the $x \times y$ dimensions (laser wavelength $\lambda_L$ of 0.8μm). The box boundary in $y$ direction is periodic. A laser pulse polarized linearly in $y$ with a Gaussian profile of $a_y = a_0 e^{-(r/\sigma_0)^2} e^{-(t-t_0)^2/\tau^2}$ traveled along $x$ axis from the left side and normally incident on the NWA target, where $a_0 = eE_0/m_e c\omega_L = 5$, $\sigma_0 = 4.3$μm and $\tau = 10\lambda_L/c = 26.7$fs. The NWA's $L$, $D$, and $S$ are set according to the achievable target parameters in the experiments. The nanowires locate in $0 < x < L$. The atomic ratio of C and H is 1:2 as in polyethylenes. A 10-μm-thick polyethylene substrate is in the region of $L < x < L + 10$μm. The initial density of electrons, protons, and C$^{2+}$ ions is 80 $n_c$, 40 $n_c$ and 20 $n_c$, respectively. Here $n_c = m_e\omega_L^2/4\pi e^2 = 1.7\times10^{21}$cm$^{-3}$ is the critical electron density, where $\omega_L$, $e$ and $m_e$ are laser frequency, electron charge and mass, respectively. The entire target is cold (10eV) and not fully ionized, and the optical field ionization is included in the setup. The number of macro-particles per cell of each ion species is 12.

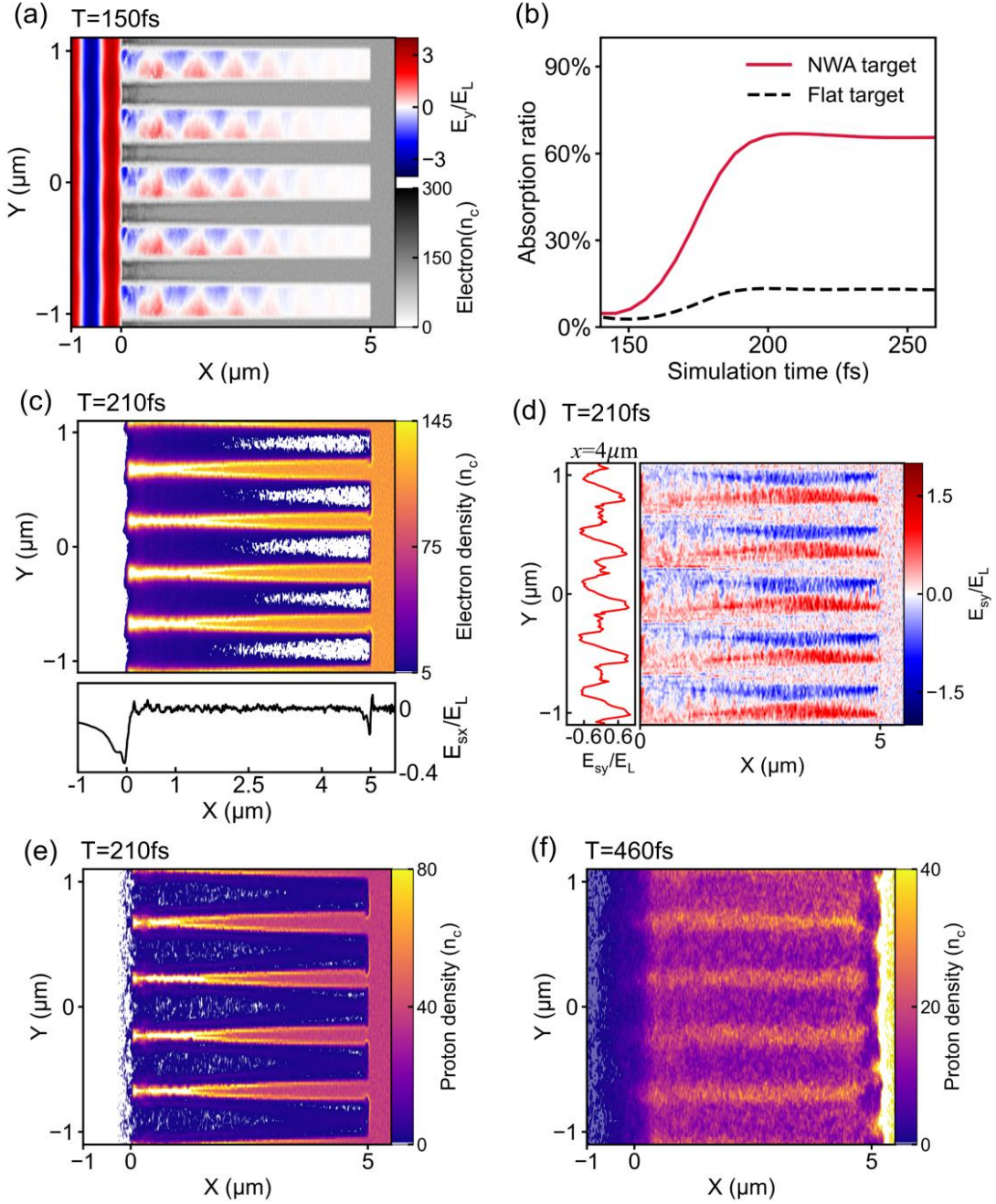

**Figure 2:** 2D PIC simulation results. (a) Spatial distribution of the electric field $E_y$ and electron density $n_e$ at t = 150 fs. (b) Laser absorption ratio over the simulation time for the planar and NWA targets. (c) Spatial distribution of the electron density at t = 210 fs, the black line shows the sheath field $E_{sx}$ along the $x$-axis at $y = 0$. (d) Spatial distribution of the sheath field $E_{sy}$ at t = 210 fs, the red lines show the sheath fields $E_{sy}$ along the $y$-axis at $x = 4\mu m$. (e) and (f) Spatial distribution of the proton densities at t = 210 fs and t = 460 fs, respectively.

Fig.2 shows the simulation results of a typical NWA target with $D = 200$ nm, $S = 450$ nm, and $L = 5\,\mu m$. At the simulation time of t = 150 fs, $E_y$ in the gaps between the nanowires (NWs) oscillates along the $x$ direction with a spatial period of $\lambda_L$. It indicates that the driving laser can propagate as deep as $5\,\mu m$ below the surface. In contrast, the laser-plasma interaction for a planar target is limited in the skin depth at the surface, and most of the laser energy is reflected. Fig.2(b) compares the absorbed laser energy for an NWA target to a planar target. The absorption ratio of 60% is five times that of the planar targets. Driven by the penetrated laser field, electrons in nanowires are drawn into the gaps and accelerated. The electron density at t = 210fs is shown in Fig.2(c). As the result of the charge separation, the sheath field is built around the NWs. The spatial distribution of the electric fields $E_{sy}$ is depicted in Fig.2(d). We make a line out of $E_{sy}$ along the $y$-axis at $x = 4\mu m$. The oscillating $E_{sy}$ which points outwards from the NWs has a maximum strength of $3 \times 10^{12}$ V/m. Ions from the NWs can gain energy in $E_{sy}$ and be laterally accelerated into gaps. For the simplicity of the discussion, we choose the protons as the representative of ions and depict their distribution in fig.2(e-f). It can be seen that the NWA evolves into a homogenous plasma with time. All the laterally accelerated protons are eventually stopped in the target, which, together with the bulk protons in the NWs, can be defined as "confined protons.". Besides the confined protons, the electrons escaped from the NWs in the $x$-direction build the field of $E_{sx}$ (at $x = 0\mu m$ in fig.2(c)), which can accelerate protons from the tip of NWs to high energy. The protons leaving the target can be called "emitted protons.". They can be found in the region of $x < 0$ in fig.2(f).

The $x - p_x$ phase diagram of protons at t = 460 fs is shown in fig.3(a), which clearly describes the characteristics of the confined and emitted protons. The emitted protons distributed at $x < 0$ have high $p_x$ up to -100 $m_e c$ (2.7MeV), while the confined protons in the range of $0 < x < 5 \mu m$ have much higher number. The $x - p_y$ phase diagram (inset) indicates the confined protons have $|p_x| < 40 \ m_e c$. The two groups of protons can be distinguished from the angular distribution in fig.3(b) as well. The emitted protons peak at 0°, while the confined protons are nearby -90° and 90°. The dashed line shows the integrated energy of protons as a function of the distribution angles in fig.3(b). It indicates that the confined protons contain energy far more than the emitted ones, contributing mostly to the plasma's proton energy density (PED). According to the above classification, the energy spectra of the two groups of protons are given in fig.3(c). The fitting line shows that the temperature of the confined and emitted protons is 50 keV and 300keV, respectively.

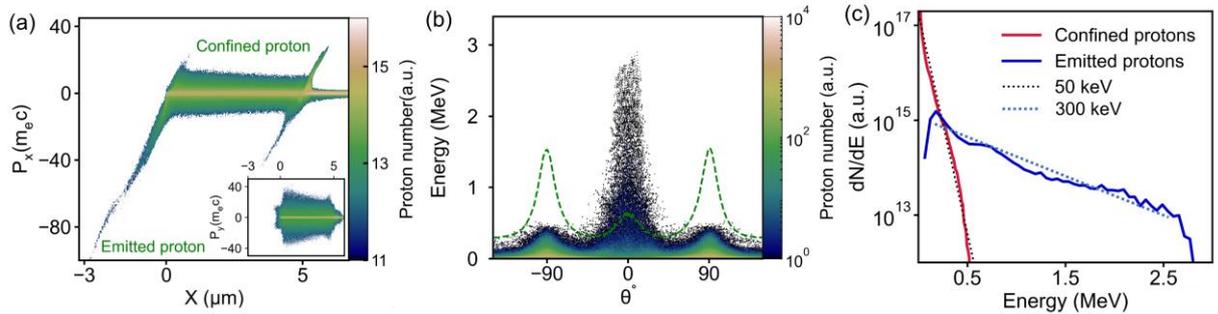

**Figure 3:** Confined and emitted protons. (a) the $x - p_x$ and $x - p_y$ phase of the protons at t = 460 fs. (d) Angular distribution of the protons at t = 460 fs. The integral energies in different angles are shown as the green line. (c) Energy spectra of confined and emitted protons at t = 460 fs. The fitted temperatures of protons are shown by the dashed lines.

# 3. Energy density of the NWA plasma

The EED and PED of a typical NWA target (the same parameters as in fig.2) and a planar target at t = 460fs are depicted in fig.4. One can see that the NWA's EED peaks at 0.3GJ/cm$^3$, about three times that of the planar target. Such an enhancement is mainly due to the higher absorption of the laser energy. Since the energetic electrons can penetrate deeply into the bulk of the target, the volume of the high-energy-density region in the case of planar is not significantly smaller than the NWA target. Compared to EED, the enhancement of the PED in the NWA target is much more prominent. The maximum PED is 0.2GJ/cm$^3$, more than one order of magnitude higher than that of planar targets. Moreover, it can be seen from fig. 4(b) that the high-PED volume in the NWA is much larger than that of the planar target. In the planar target, the depth of the high-PED region is limited to 1 μm below the surface. While in the NWA target, the high-PED region covers the whole target. This is because the laterally accelerated protons widely exist in the NWA, carrying a significant amount of energy.

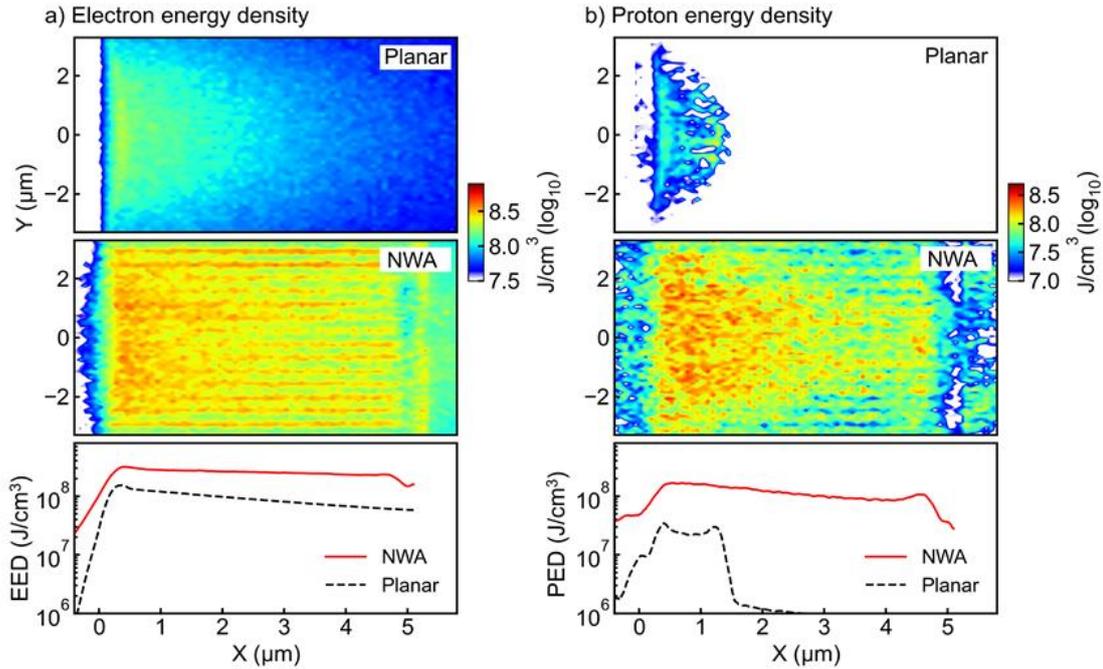

**Figure 4:** Spatial distribution and the line out (*y* = 0) of (a) EED and (b) PED for a planar target and a typical NWA target at the simulation time of t = 460 fs.

To reveal the dependence of PED on the target parameters of NWA, we perform a series of simulations. Fig.5(a-d) shows the spatial distribution of PED of NWA targets with *L* = 5, 2, 1 μm (*D* = 200 nm, *S* = 450 nm), and *L* = 5 μm (*D* = 300 nm, *S* = 450 nm), respectively. Fig.5(e-f) depicts the line out of the PED along the *x*- and the *y*-direction, respectively, where the data from an NWA of *L* = 10 μm, *D* = 200 nm, and *S* = 450 nm is also included. It can be found that NWA targets with different lengths have almost the same maximum PED. Increasing the length of NWs does not enhance the PED but would enlarge the volume of the high-PED region. In a given NWA, the PED decreases with $x$. For the $L = 10$ μm NWA, the PED drops to 0.05 GJ/cm$^3$ at $x = 10$ μm, which is about 1/4 of the peak value at the surface. The decreasing gradient strongly depends on the $D$. The $D = 300$ nm NWA has a gradient of two times that of $D = 200$ nm NWA. So the total proton energy contained in the plasma is significantly lower as a result.

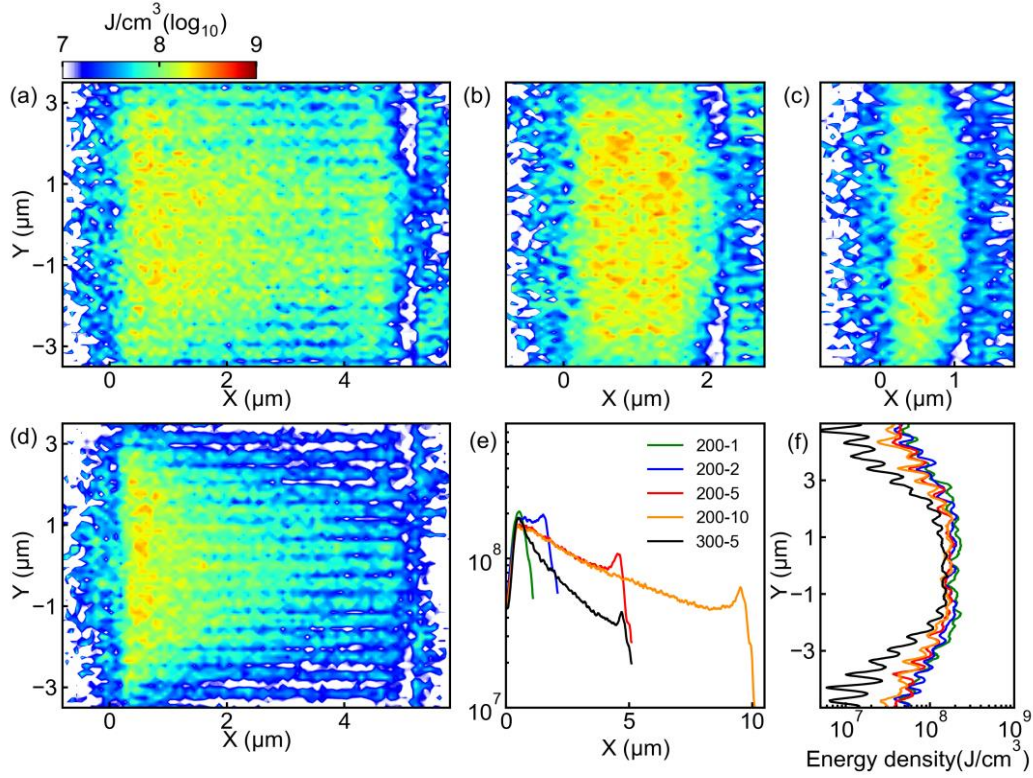

**Figure 5:** Spatial distribution of PED for NWA targets with (a) $L = 5$μm, (b) $L = 2$μm, (c) $L = 1$μm ($D = 200$nm, $S = 450$nm) and (d) $L = 5$μm ($D = 300$nm, $S = 450$nm) respectively. (e) (f) PED for different targets along the $x$ and $y$ direction, respectively.

Fig.6 gives the laser absorption ratio, temperature and number of the confined protons, and the total energy of emitted and confined protons as a function of the NW's diameters or lengths. It should be noted that the confined protons with energy lower than 1keV are not counted there. As shown in fig.6(a), the laser absorption ratio is closely related to the NWs' diameters—smaller diameters correspond to higher absorption. This is because a larger diameter means narrower gaps. The time before the complete filling of the gaps by overdense plasma is, therefore, shorter, resulting in lower laser absorption and incompleted heating of NWs. The temperatures and numbers of >1keV confined protons therefore decrease with the increasing diameter. On the other hand, the decreased diameters will reduce the average density of the targets for a given $S$, which is unfavorable to higher proton numbers. Fig.6(b) shows that a moderate diameter of $D = 200$nm is the best tradeoff, where the average density is 16% of solid density, and energetic confined protons contain 63 mJ energy for the 1J input laser energy.

At a fixed diameter of 200nm, we change the lengths of NWs to study their influence on the NWA's PED. Fig.6(c) indicates that increasing nanowires' lengths will not affect the temperature of confined protons and the laser absorption. This is understandable as they are determined by the laser-plasma interaction at the sidewall of the NWs. In contrast, the number of the confined protons is proportional to the length of the NWs. Moreover, as shown in fig. 6(d), with the increase of the length of the NWs, the energy converted to confined protons rises, while that to emitted protons drops. This is consistent with the observation in fig.5 that NWA targets with longer NWs have larger high-PED volumes. It can be seen that 9% of the laser energy is converted to protons in the best case ($L = 10\mu m$), and the vast majority(>90%) of which is to the confined protons.

In summary, our simulations explore the parameter space of NWA for the purpose of a high PED. It turns out that the optimal NW diameter is 200nm. With the increase of the length of NWs, the energy to confined protons rises, whereas it declines to emitted protons. NWA targets with a length of 10µm and a diameter of 200nm perform best, where 8% of laser energy can be converted to confined protons

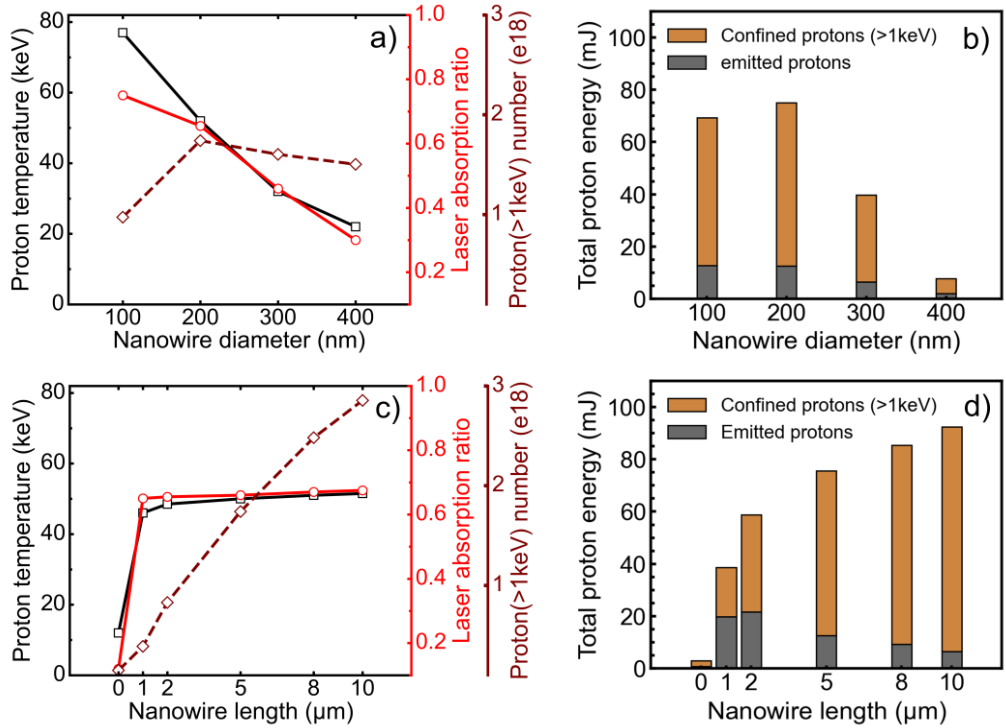

**Figure 6:** (a) laser absorption ratio, confined proton temperature, confined proton number (>1keV), and (b) the energy proportion of emitted and confined protons as a function of NWs' diameters. (c)(d) as a function of NWs' lengths. The other parameters of NWA targets are $S = 450$nm, $L = 5\mu m$ in (a)(b) and $D = 200$nm, $S = 450$nm in (c)(d).

# 4. Experimental campaign

Guided by the simulation results, we performed an experimental campaign to study the high-energy-density states in NWA targets. Confined ions are challenging to be detected in the experiment. Instead, we measured the emitted ions to verify our conclusion.

The experiment was conducted on the CLAPA laser facility with a central wavelength of 800 nm and a duration of 30 fs (full width at half maximum, FWHM) at Peking University. Fig.7(a) depicts the schematic of the experimental setup. A linearly polarized laser pulse was normally irradiated on the NWA target. In order to prevent the pre-expansion of nanowires, a plasma mirror system was employed to improve the laser-contrast ratio, ensuring the intensity of pre-pulse is below $10^{11}$W/cm$^2$ at 5ps(53). We checked whether the NWs could maintain their structure under the irradiation of the ps-pedestal before the main pulse by employing hydrodynamic simulation code MULTI2D(54). The results indicate the scale length of the pre-plasma $L_{pre} = n_e/(dn_e/dx) \sim 22$nm, insignificant compared to the gaps between nanowires. The on-target laser energy was about 1J. Ultimately, the laser beam was focused into a spot size of 4.0-4.4 $\mu$m (FWHM) to reach an on-target peak intensity of $4.8\times10^{19}$ W/cm$^2$ ($a_0 \sim 5$) by using an f/3, 90° and gold-coated off-axis-parabolic mirror (OAP). The OAP has a 3 mm diameter hole that enables ions from the front side of the targets to pass to the ion spectrometer.

The ions spectra were measured using a Thomson Parabola Spectrometer (TPS) in the direction normal to the front side of the target. The TPS was composed of a collimating aperture, magnets, electrodes, a microchannel plate, a phosphor screen, and an electron-multiplying CCD(EMCCD). After the 1mm-wide collimating aperture ($1.3\times10^{-6}$ Sr), the collimated ions with different energy and charge to mass ratio (CMR) are deflected by the electromagnetic fields and hit different positions of the microchannel plate. The ion signal multiplied by the microchannel plate was converted to the optical signal at the phosphor screen and captured by the EMCCD. The ions with different CMR have their own parabola traces imaging in TPS raw data. Besides, the angular distribution of emitted protons was recorded by CR39 ion track detectors placed at the angle of 20°, 40°, 55°, and 80° to the target normal.

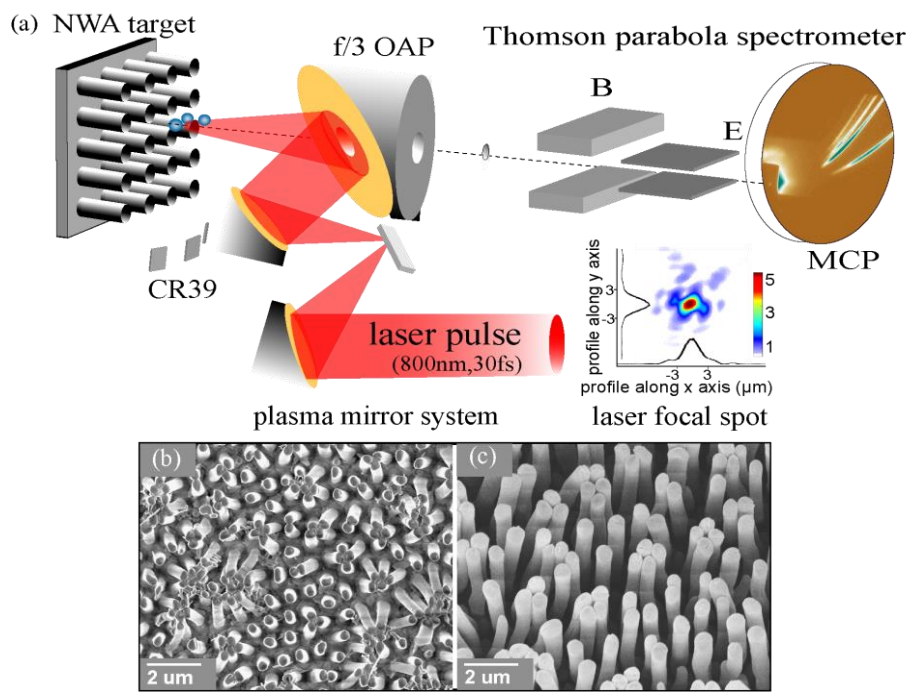

**Figure 7:** (a) Schematic of the experimental setup. SEM image: (b) Top and (c) Side view of the NWA target.

**Table 1:** Parameter of all NWA targets and planar targets.

| Target | $D$ (nm) | $S$ (nm) | $L$ (μm) | $\rho_{av}/\rho_s$ | Thickness (μm) |
|---|---|---|---|---|---|
| | 200 | 450 | 1,2,4,5,10 | 0.16 | |
| **NWA** | 300 | 450 | 5 | 0.35 | 450±50 |
| | 500/600 | 800 | 5 | 0.31/0.45 | |
| **Planar** | --- | --- | --- | --- | 450±50 |

Polyethylene NWA targets with a variety of parameters were shot in the experimental campaign. They were prepared by the following procedure. Firstly, a thick polyethylene sheet was attached to an Anodized Aluminum Oxide template (AAO)(55), where nanochannels were hexagonally packed. By heating and mechanically compressing the sheet, the polyethylene molecules were stimulated into the templates. Afterward, an NWA target was obtained by dissolving the AAO membrane in a 2M NaOH solution for 30 minutes. The diameters, spacings, and lengths of the NWs were determined by the AAO templates. The morphology of NWAs was characterized by scanning electron microscopy (SEM), and its top and side views are shown in fig.7(b-c). The thickness of the polyethylene substrate that supports the nanowires is about $450\mu m$. The average density of the NWAs lies in the range of 16 ~ 45% intrinsic solid density. Planar polyethylene targets with a thickness of 450 $\mu m$ were shot for comparison. Table 1 summarizes the parameters of all NWA targets and planar targets.

## 4.1 Measurements of the emitted protons

The experimental results of emitted protons for NWA targets with different lengths and diameters are summarized in fig.8(a-b), where the error bars reflect the shot-to-shot fluctuations and the dots are the arithmetic means. The maximum proton energies of planar targets are depicted by the zero-diameter and zero-length points. The solid lines represent the 2D PIC simulation results, which fit the experimental results. The corresponding best experimental proton spectra are displayed in each figure respectively [see fig.8(c-d)]. Compared to the planar targets, a remarkable enhancement of maximum energy for NWA targets can be observed.

The dependence of the nanowires' diameters on the maximum proton energy can be obtained from fig.8(a) and (c). In both series of $S = 450$ nm and $S = 800$ nm, the maximum proton energies decline with increasing NW's diameters, which fits the simulation results well. For NWAs with fixed $D = 200$ nm and $S = 450$ nm, the maximum proton energies rise with the increased length until $L = 2$ μm, then decline. Such a trend reflects that the strongest $E_{sx}$ is produced in the laser-plasma interaction for the 2-μm-length NWAs. When the NWs' length exceeds the optimal length, the laser pulse propagates deeper into the NWA and transfers more of its energy to confined protons. It should be noted that our experimental results about the emitted are consistent with that in another recent report(56).

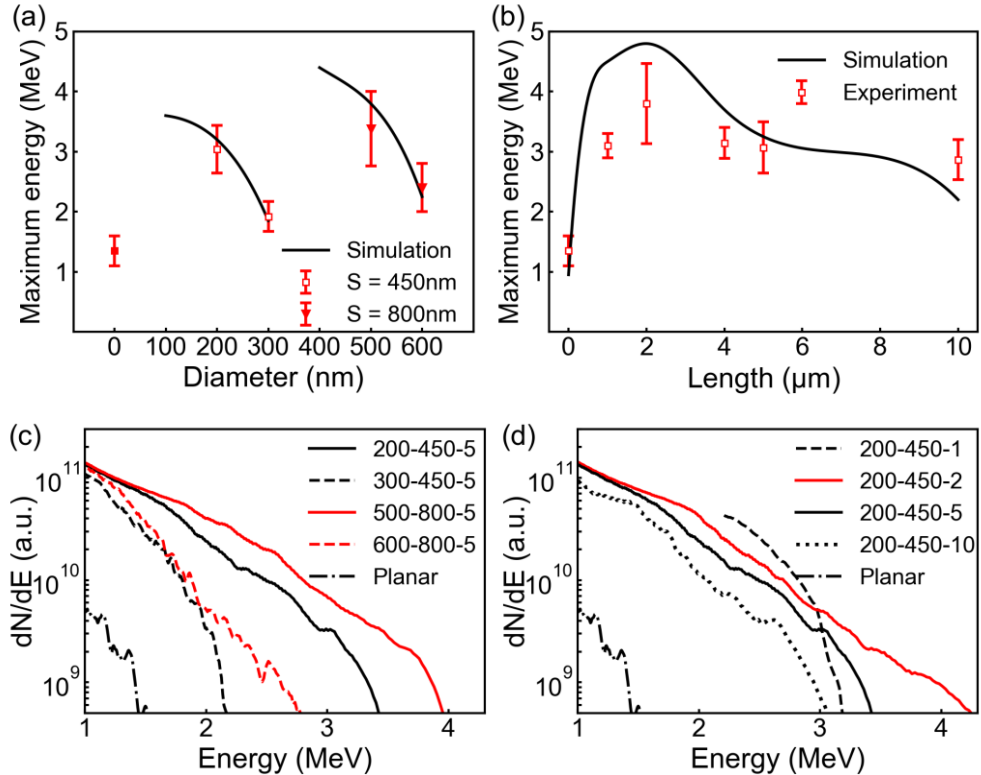

**Figure 8:** Experimental and simulated maximum proton energy results for varying (a) NWA's diameters and (b) lengths. The other parameters of NWA are $S = 450$nm / $S = 800$nm, $L = 5$μm in (a) and $D = 200$nm,

$S$ = 450nm in (b). (c) and (d) depicts the proton spectra in (a) and (b), where the NWA target with $D$ = 200nm, $S$ = 450nm, and $L$ = 5μm is abbreviated as 200-450-5.

The angular distribution of emitted protons was measured by the CR39 after 16 shots in a row. 30-μm-thick aluminum foils were placed on the CR39, which can block most carbon ions below 32MeV and protons below 1.6MeV. The images of etched CR39 placed at 20°, 55°, and 80° to the target normal are displayed in fig.9(b-d). They were etched for 12 h in 6N NaOH solution at 70°C after shooting. The sizes of the three CR39 plates are identical, corresponding to a solid angle of $2\times10^{-6}$ Sr. As one can see in fig.9(d), only a few protons were emitted at 80° with respect to the target normal. More proton tracks were observed with the decrease of the angle. The proton tacks (the small dots in the image) at 20° are very dense, even close to saturation. At the angle of 55°, a number of energetic proton tracks were observed, which implies that the divergence of the protons from the NWA is larger than typical planar targets. In general, the CR39 result is inconsistent with our simulation results in fig.3(b). The most energetic confined ions in lateral directions in fig.3(b) cannot be detected as they had been stopped in the targets.

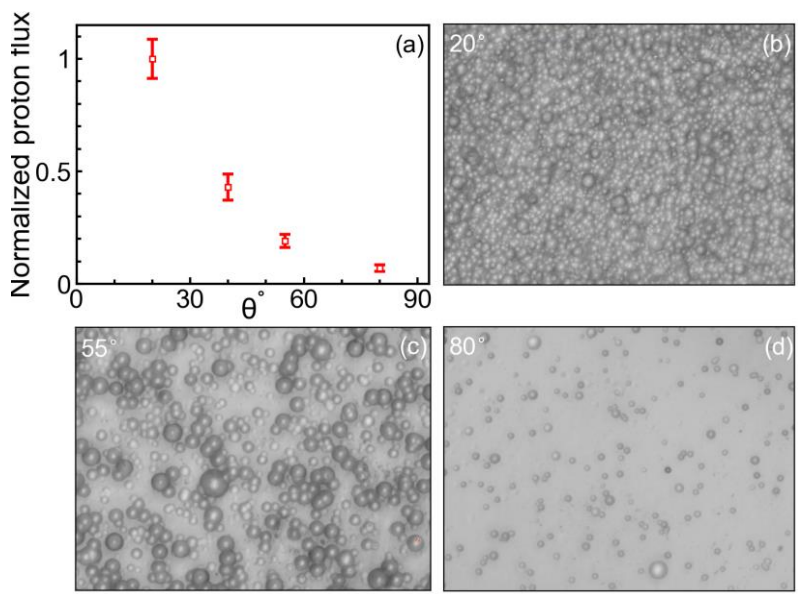

**Figure 9:** (a) Angular distribution of the protons. The error bars represent the standard deviation of the statistics. (b-d) are the raw data of CR39 with the same size at the angle of 20°, 55°, and 80° with respect to the target normal.

## 4.2 Deuteriums and neutrons generated from deuterated polyethylene NWA

An NWA absorbs a majority of the laser energy and forms a near-solid-density HEDP, which creates an excellent condition for charged-particle nuclear reactions. In the experimental campaign, we also employed deuterated polyethylene NWA targets to demonstrate the neutron generation from the $^2$H(d, n)$^3$He fusion. They were prepared following the same procedure as the polyethylene NWAs.

We first tried to measure the deuterium ion (D$^+$) emitted from NWA targets utilizing the TPS. The raw data, including all ions with different CMRs, is presented in fig.10(a). The traces with CMRs of 1, 1/2, ... can be identified. However, the traces of D$^+$ and C$^{6+}$ are indistinguishable because they have the same CMR of 1/2. Therefore, we positioned a 20-μm-thick Al foil in front of the MCP to distinguish them. Due to different stopping powers, C$^{6+}$ with energy < 1.83 MeV/u will be blocked by the foil, while D$^+$ with energy > 0.78 MeV/u can get through [see fig.10(c)]. It allows us to identify the D$^+$ in the range of 0.78-1.83 MeV/u on the MCP. Fig.10(b) displays the raw TPS data. The broadening of traces is due to the straggling when ions pass the Al foil. The cut-off of the trace at the high-energy end corresponds to 0.8 MeV/u, which is very close to the theoretical value of 0.78 MeV/u. It's unambiguous that the remaining parabola trace of CMR = 1/2 only results from the D$^+$.

By analyzing the TPS traces, the energy spectra of ions are obtained and shown in fig.10(d-e). The maximum proton energy of 3.5 MeV is similar to the one without the Al foil, which indicates the good repeatability of the results. The maximum energy of the deuterium ions is 1.5 MeV/u, significantly lower than that of the protons. No deuterium ions signal was detected for planar deuterated polyethylene targets.

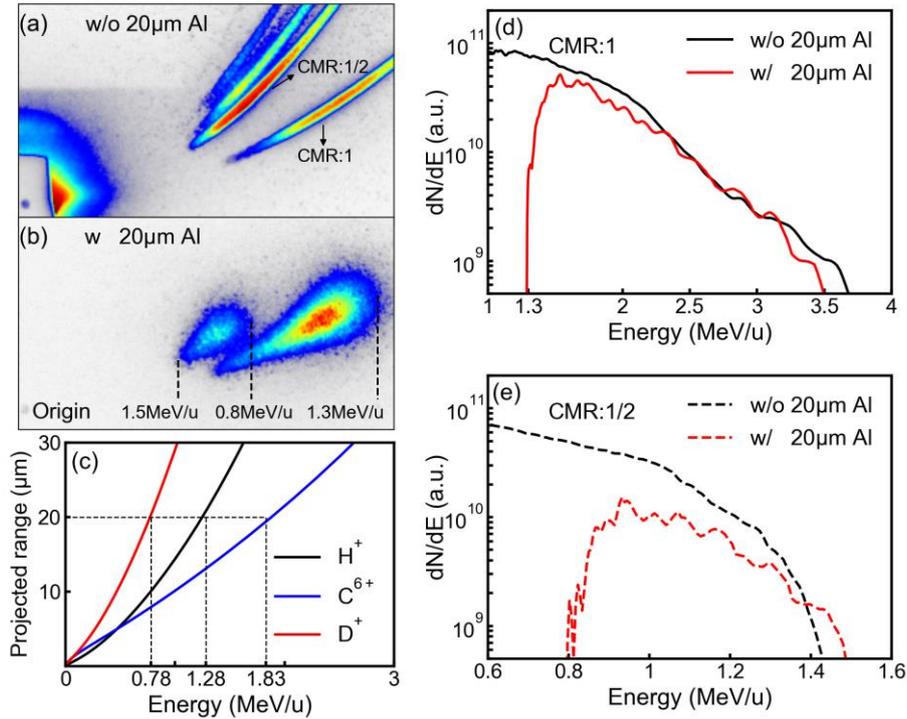

**Figure 10:** Ion spectra: TPS raw data (a) without and (b) with 20μm aluminum foil. (c) Projected range in aluminum foil for proton, deuterium, and carbon ions. The energy spectra of ions with (d) CMR = 1 (proton) and (e) CMR = 1/2 (deuterium or carbon ion).

The observed emission of MeV-level deuteriums is a signature of the strong laser-plasma coupling in the NWAs. Similar to the case of polyethylene NWA, we performed 2D simulations to study the HEDP in the deuterated polyethylene NWA by employing the SMILEI code(57). The advantage of SMILEI over EPOCH is that it includes a nuclear reaction module, which can seamlessly simulate the $^2$H(d, n)$^3$He fusion reaction with the Monte Carlo method. The parameters of the simulation are the same as that before.

Fig.11(a) shows the simulated energy density of deuterium ions in the deuterated NWA target with $D = 200$nm, $S = 450$nm, and $L = 5\mu$m at t = 460 fs. The distribution of the energy density is very uniform, with a peak value of 0.22 GJ/cm$^3$. The peak temperature of confined deuterium ions is 80keV. As the products of the fusion reactions, the distribution of helium ions is depicted in fig.11(b). It indicates that nuclear reactions mainly happen in the focal spot area where the deuterium ions' temperature is highest. Fig.11(c-d) shows the total helium yield (dashed line) from NWAs with different lengths and diameters at t = 3ps. Compared to the planar target, more than 20 times higher yields can be obtained for the optimal target. Increasing NWA's length leads to the enhancement of yield, which, however, becomes less prominent when $L$ exceeds 5 μm. The optimal diameter is $D = 200$nm. The corresponding density is about 16% solid density.

The D-D nuclear reaction produces the same number of helium ions and neutrons. In recent work, we reported the measurement of the neutrons from NWA targets with $D = 200$-500nm, $S = 450$-800nm, and $L = 1$-10 μm(58), where the average neutron yield per shot was $(24 \pm 18) \times 10^6$/J. One can see that the simulation results here agree with our previous observation well.

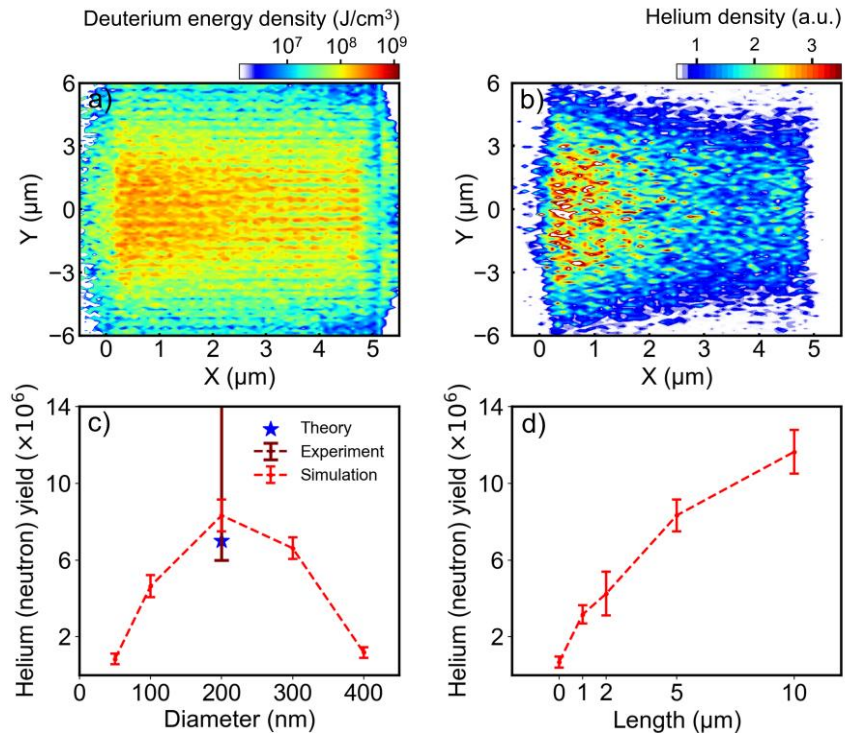

**Figure 11:** (a) Deuterium energy density at the simulation time of t = 460 fs. (b) Spatial distribution of the density of helium ions at t = 460 fs. Simulated helium (neutron) yield for different (c) NWs' lengths and

(d) diameters at the simulation time of t = 3ps. The error bar represents the statistical fluctuations. The theoretical value of neutron yield from the NWA target with $D = 200$nm, $S = 450$nm, and $L = 5\mu$m, and the average neutron yield in the experiment from different NWA parameters are also shown in (c).

We can also theoretically estimate the neutron yield. The volumetric reaction rate is given by $R = n_D^2 \sigma \vartheta_r / (1 + \delta)$. $n_D$ is the average deuterium ion density of $16\% \times 8 \times 10^{22}$ cm$^{-3}$ for the NWA target with $D = 200$nm, $S = 450$nm, and $L = 5\mu$m. $\vartheta_r$ is deuterium-deuterium relative velocity, which we take as $2\vartheta_D \sim c/54$ based on the deuterium ion temperature of 80keV. $\sigma$ is the corresponding fusion cross-section of 0.085barn(59) for deuterium ion energy of 160keV in the center-of-mass frame. $\delta$ is the Kronecker symbol of 1. Therefore, the reaction numbers can be roughly estimated as $N = V n_D^2 \sigma \vartheta_r \tau / 2$, multiplied by the reaction volume $V$ and the reaction time $\tau$. The $V$ is about $6 \times 10^{-10}$ cm$^{-3}$ considering the cylindrical reaction area of $6\mu$m radius and $5\mu$m length. $\tau$ can be roughly estimated by $V^{1/3}/\vartheta_D \sim 2.8$ps. According to the above analysis, the number of neutrons $N \sim 7 \times 10^6$ in $4\pi$, which is in agreement with the experimental and simulation results.

# 5. Conclusion

In conclusion, we numerically and experimentally studied the HEDP formed in femtosecond-laser-irradiated NWA targets. It's found that the unique structure of the NWs leads to the generation of rich, energetic ions, which eventually are confined in the plasma and contribute most to the high energy density state. The emitted ions, on the other hand, can be used as valuable diagnostics of the interaction. Our systematic study on the dependence of the ion-energy-density on the parameters of the NWA targets illustrates the optimal parameters for the HEDP in NWAs and gives a valuable guide for future works that employ NWA targets. The achieved large-volume homogenous GJ/cm$^3$-level energy density of ions driven by merely J-level lasers would provide a valuable testbed for high-energy-density physics and nuclear physics.

# Acknowledgments

This work was supported by the NSFC innovation group project (11921006), the National Grand Instrument Project (2019YFF01014402), United States Department of Energy under Grant No. DE-FG03-93ER40773, and NNSA Grant No. DENA0003841 (CENTAUR). The PIC simulations were carried out in the High-Performance Computing Platform of Peking University.